# Crystalline Field Effects on Magnetic and Thermodynamic properties of a Ferrimagnetic Centered Rectangular Structure

Maen Gharaibeh*, Rama Abu Haifa, Abdalla Obeidat, Mohammad H.A. Badarneh, M.K Qaseer

Department of Physics, Jordan University of Science and Technology, Irbid 22110, Jordan

*Corresponding author: magh@just.edu.jo

**Abstract**

The magnetic properties and phase diagrams of the mixed spin Ising model, with spins $S = 1$ and $\sigma = \frac{1}{2}$ on a centered rectangular structure, have been investigated using Monte Carlo simulations based on the Metropolis algorithm. Every spin at one lattice site has four nearest-neighbor spins of the same type and four of the other type. We have assumed ferromagnetic interaction between the same spins type, antiferromagnetic for different spin types. An additional single-site crystal field term on the $S = 1$ site was considered. We have shown that the crystal field enhances the existence of the compensation behavior of the system. In addition, the effects of the crystal field and exchange coupling on the magnetic properties and phase diagrams of the system have been studied. Finally, the magnetic hysteresis cycles of the system for several values of the crystal field have been found.

**Introduction**

The topic of mixed spins with different orders is the subject of many areas in physics. For instance, the mixed-spin magnetic systems have been studied vigorously in the last few years by applying different models such as the Heisenberg model [1], XY-model [2], and Ising model [3], to predict their magnetic and thermodynamic properties. In particular, the mixed-spin Ising model has been used as the simplest model of ferrimagnetic materials. The most important property of ferrimagnetic materials is the appearance of the so-called compensation temperature(s). These non-critical points have zero total magnetization. In contrast, sublattices of the system retain their magnetization against extrinsic effects such as temperature [4]. This phenomenon in which the total magnetization is zero can be observed experimentally by tuning the temperature below the critical temperature. Such compensation points have many technological applications such as thermo-magnetic recording media, dynamic and random-access memories and magneto-optical recording media devices [5-9]. Theoretically, many methods have been used to study these systems based on statistical mechanics such as renormalization group [10-12], low and high-temperature series expansion [13-16], finite cluster approximation [17], mean-field theory [18], effective field theory [19, 20], and Monte-Carlo simulations [21-24].

The Monte-Carlo simulations might be the most reliable technique to study different real/ (artificial) structures in two and three dimensions of ferrimagnetic materials with different exchange interactions and different spins. The crystal field effect on magnetic properties has been studied by many researchers with different types of spins [25-32] but has still not been investigated thoroughly in many important structures

This paper extended our previous work [24] to include the effects of single-ion anisotropy and applied magnetic fields on magnetization behaviors and compensation points. We will apply Monte Carlo techniques to study the different magnetic properties of the ferrimagnetic spin-1– spin-1/2 magnetic atoms arranged in a 2-D centered rectangular structure. The structure consists of two sublattices: sublattice A of spin $S = 1$ and sublattice B of spin $\sigma = 1/2$. Each sublattice has a rectangular structure.

The centered rectangular structure consisting of two sublattices with different coupling strengths (in our case, five different coupling strengths) might be one of the most important structures; since this structure can be looked at differently. For example, instead of looking at the structure as a two-dimensional structure, one might look at it as a bilayer structure with one layer consisting of one type of atom arranged in a rectangular sheet. The vertex of a second similar structure layer lies above the center of the first and has a different type of atom. Adding extra bilayers to the structure allows us to study different structures such as bcc and tetragonal structures. Currently, we are in the process of calculating the coupling strengths of FePt, FeNi, FePd and of CoPt, NiPt, $CoNiPt_2$, and of $FePt_3$, $CoPt_3$, $NiPt_3$, and of $Fe_3Pt$, $Ni_3Pt$, $Co_3Pt$ using the density functional theory, then applying the Monte-Carlo simulations to study different thermodynamic and magnetic properties of these structures.

## 2. Model and formalism

### 2.1 Lattice structure and Hamiltonian

The rectangular structure adopted here is the same as in [24]. It consists of two sublattices-*A* and *B*; sublattice-*A* is composed of *S*-type of spins-1, whereas sublattice-*B* is composed of *σ*-type of spins-1/2 (see Fig. 1). The coordination number of each site is 8. The Hamiltonian of the magnetic structure includes nearest neighbor interactions; the external magnetic field and the crystal field is given as:

$$H = -J_1 \sum_{\langle i,j \rangle} S_i S_j - J_2 \sum_{\langle l,m \rangle} \sigma_l \sigma_m - J_3 \sum_{\langle i,k \rangle} S_i S_k - J_4 \sum_{\langle l,n \rangle} \sigma_l \sigma_n - J_{12} \sum_{i,l} S_i \sigma_l \quad (1)$$
$$- B \sum_{i,l} (S_i + \sigma_l) - H'$$

where $H'$ is defined as follows:

$$H' = D \sum_i S_i^2 \qquad (2)$$

$H'$ represents the crystal field interaction with sublattice-A of spins-1. In our simulations, the spin moments of each spin are $S = 1$ and $\sigma = \frac{1}{2}$. Hence, we associate the $(2S + 1)$ and $(2\sigma + 1)$ possible spin projections $(-1, 0, 1)$ and $\left(-\frac{1}{2}, \frac{1}{2}\right)$, respectively. The other possible five interactions in $H$ are, namely, along [10] direction, $J_1$ is the $S - S$ exchange interaction, $J_2$ is the $\sigma - \sigma$ exchange interaction. Along [01] direction, $J_3$ is the $S - S$ interaction, $J_4$ is the $\sigma - \sigma$ interaction. Diagonally $J_{12}$ is the $S - \sigma$ ferrimagnetic interaction. The summation indices $<i, j>$ and $<l, m>$ denote the summations over all nearest-neighbor spins $S - S$ and $\sigma - \sigma$, respectively. On the other hand, the summation indices $<i, k>$ and $<l, n>$ denote the summations over all nearest-neighbor spins $S - S$ and $\sigma - \sigma$ along [01] direction, respectively. We have taken $J_1$, $J_2, J_3$, and $J_4$ positive to ensure ferromagnetic interaction and $J_{12}$ negative for antiferromagnetism. $D$ can have positive or negative values. In this work, the different values of the exchange parameters are kept the same for all simulations presented here. Those parameters are chosen such that $T_{comp}$ and $T_c$ exist and taken from our previous work where the crystalline fields effect are ignored [24].

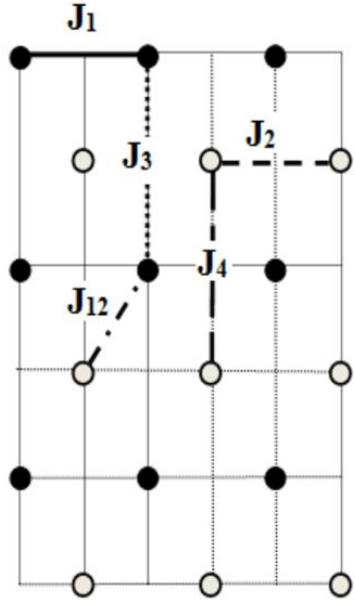

**Fig. 1.** Schematic representation of the centered rectangular structure formed by sublattice-A and B with spin-1 and spin-$\frac{1}{2}$, respectively. The filled circles represent magnetic atoms (spin $S = \pm 1, 0$) in sublattice-A, whereas the open circles represent magnetic atoms (spin $\sigma = \pm\frac{1}{2}$) in sublattice-B.

## 2.2 Monte Carlo Simulation and Calculations

We implement the Monte Carlo simulation technique based on the metropolis algorithm [33] and apply periodic boundary conditions in the $x$ and $y$ directions. To generate new configurations, we choose a spin at random and then flip it. By making spin-flip attempts, each flip is accepted or rejected according to the metropolis algorithm. Data were generated using 1000000 Monte Carlo simulations to equilibrate the system, followed by 850000 Monte Carlo steps for each spin configuration. The total number of the spins in our simulation is $N_{tot} = 2450$, which contains $N_A = 35 \times 35$ spins of the $S$-type in sublattice-$A$ and $N_B = 35 \times 35$ spins of the $\sigma$-type in sublattice-$B$. Several additional simulations were performed for number of atoms in one direction $L = 50$, and $100$, but no significant differences were found from the results presented here. Calculation of the error is based on the method of blocks; the $L$ size is divided into $n_b$ blocks of length $L_b = L/n_b$. The number of blocks is chosen such that $L_b$ is higher than the correlation length. Therefore, error bars are calculated by grouping all the blocks, then taking the standard deviation [34]. In this paper, the temperature of the system is measured in units of energy. Our program calculates the following parameters, namely:

The magnetization per site of sublattice-$A$ and $B$ can be calculated by

$$M_A = \frac{1}{N_A} \left\langle \left| \sum_{i=1}^{N_A} S_i \right| \right\rangle \tag{3}$$

$$M_B = \frac{1}{N_B} \left\langle \left| \sum_{l=1}^{N_B} \sigma_l \right| \right\rangle \tag{4}$$

$$M_{tot} = \frac{N_A M_A + N_B M_B}{N_A + N_B} \tag{5}$$

The sublattice magnetic susceptibilities are given by

$$\chi_A = N_A \beta (\langle M_A^2 \rangle - \langle M_A \rangle^2) \tag{6}$$

$$\chi_B = N_B \beta (\langle M_B^2 \rangle - \langle M_B \rangle^2) \tag{7}$$

$$\chi_{tot} = \frac{N_A \chi_A + N_B \chi_B}{N_A + N_B} \tag{8}$$

where $\beta = \frac{1}{k_B T}$, $T$ is the absolute temperature, and $k_B$ is the Boltzmann factor, assumed to its unit value $k_B = 1$, for all our numerical calculations.

Finally, we have calculated the specific heat of the system as follows:

$$\frac{C_v}{k_B} = \frac{\beta^2}{N_{tot}} (\langle H^2 \rangle - \langle H \rangle^2) \tag{9}$$

At the compensation temperature, $T_{comp}$, the sublattice magnetizations cancel each other, and the total magnetization of the system is zero. Hence, to determine $T_{comp}$ from the computed magnetization data, a crossing point of the absolute value of sublattice $A$ and $B$ magnetizations needs to be determined under the following condition:

$$|M_A(T_{comp})| = |M_B(T_{comp})| \tag{10}$$

$$sign\left(M_A(T_{comp})\right) = -sign\left(M_B(T_{comp})\right) \tag{11}$$

with $T_{comp} < T_C$ where $T_C$ is the critical temperature. In this paper, $T_C$ is determined from the divergence of the susceptibilities curves.

## 3. Results and Discussions

This section will present our results of the magnetic and thermodynamic properties of the mixed-spin ferrimagnetic centered rectangular structure obtained using the Monte Carlo simulations. We have observed the influence of Hamiltonian parameters on the phase diagrams, magnetization, susceptibility, and specific heat of the system, and finally obtained hysteresis loops. Note that the error bars are smaller than the point markers for all figures and do not show up in the figures.

### 3.1. *Phase diagrams*

Fig. 2 illustrates the influence of the exchange coupling $J_1$ on the critical and compensation temperatures for different values of the crystalline field, keeping $J_2, J_3, J_4, J_{12}$ and $B$ constant. The figure shows a gradual increase in the critical temperature value. While the compensation temperature increases linearly as the value of the exchange coupling $J_1$ increases. Furthermore, we have noticed that for any value of the exchange coupling $J_1$ the compensation temperature of the system increases as the crystalline field increases. That is mean more atoms of sublattice-$A$ prefer to stay in spin -1 state for positive values of $D$ and in spin +1 state for negative values of $D$ as the temperature increases. It is worth mentioning that for $J_1 < 0.25$ the compensation temperature appears only for the easy axis anisotropy case. Furthermore, for $J_1 \geq 2.5$, the compensation temperature appears only for the hard axis anisotropy case.

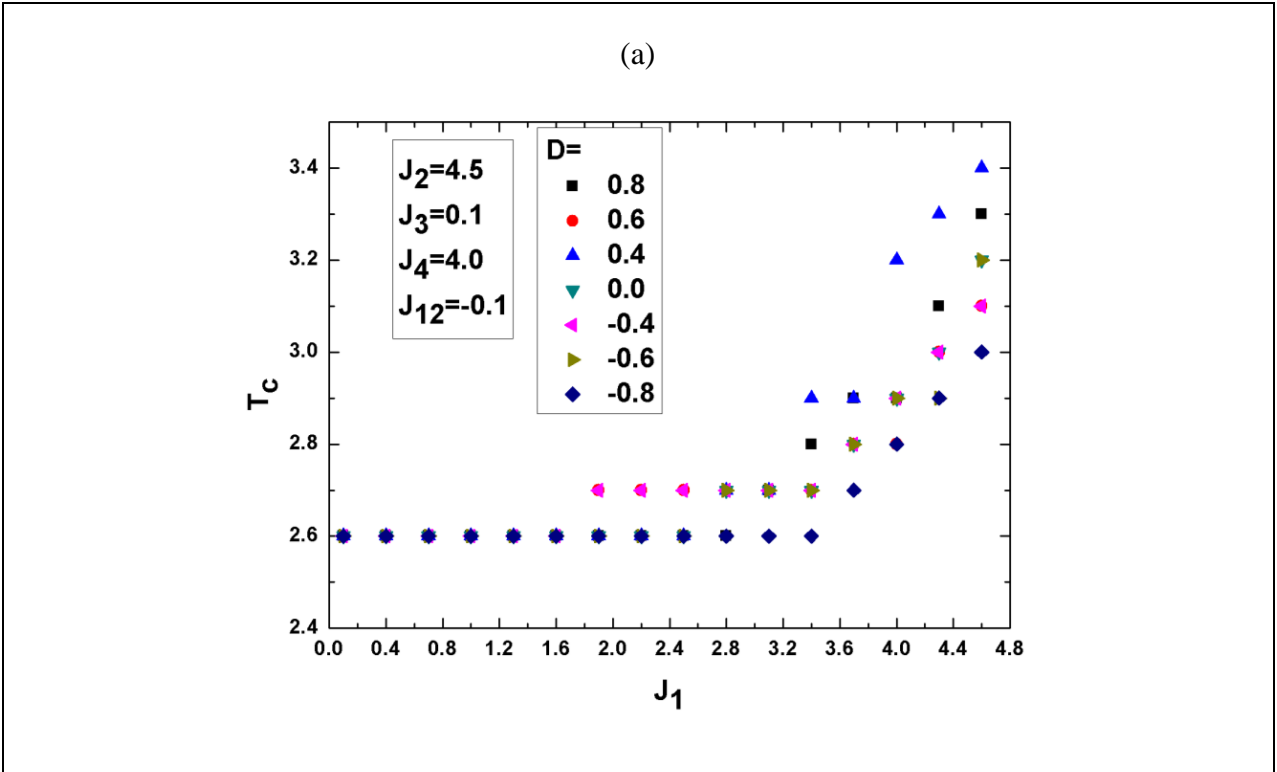

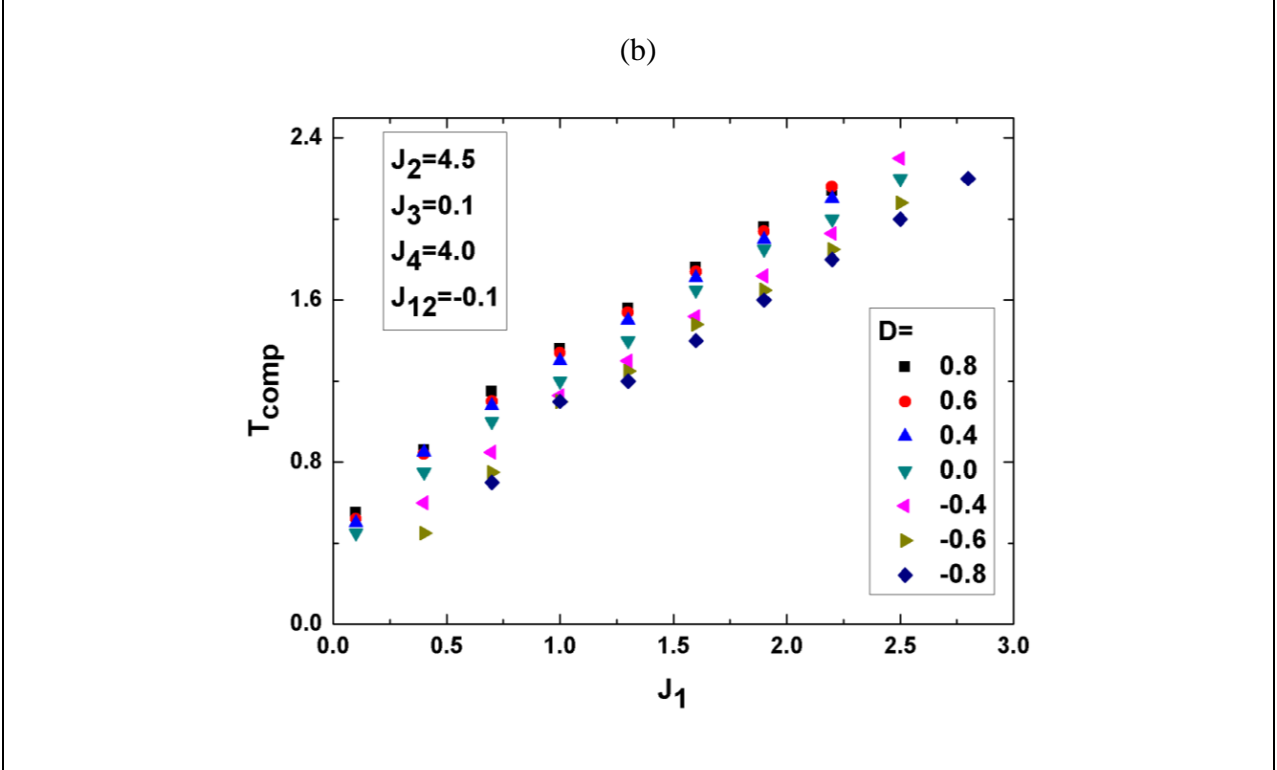

**Fig. 2.** The phase diagram (a) $(T_C, J_1)$ and (b) $(T_{comp}, J_1)$ at $J_2 = 4.5$, $J_3 = 0.1, J_4 = 4.0, J_{12} = -0.1, B = 0$ and for different values of the crystalline field.

To explore the effect of the crystalline field $D$ on the critical and compensation temperatures, we have plotted in Fig. 3 the phase diagram of the system at $J_1 = 0.3$, $J_2 = 4.5$, $J_3 = 0.1$, $J_4 = 4.0$, $J_{12} = -0.1$, and for different values of the crystalline field in the absence of the external magnetic field. The critical temperature of the system stays constant with varying the crystal field. The figure shows that for $D = 0$, the system exhibits critical and compensation temperatures, which can also be confirmed from Fig. 4. On the other hand, when decreasing the negative value of the crystal field, the compensation temperature decreases. This is because introducing a hard axis anisotropy in the system means a very low temperature is needed to ensure a perfectly ordered sublattice-$A$. Hence, the crossing point of the absolute value of sublattice-$A$ and $B$ magnetizations decreases, thereby the compensation temperature of the system decreases. Increasing the positive value of the crystalline field results in increasing the system's compensation temperature, which is due to introducing the easy axis anisotropy in the system, suggesting that a higher temperature is needed to disorder sublattice-$A$ by forming magnetic domains inside it. Hence, the crossing point of the absolute value of sublattice-$A$ and $B$ magnetizations increases, increasing the system's compensation temperature.

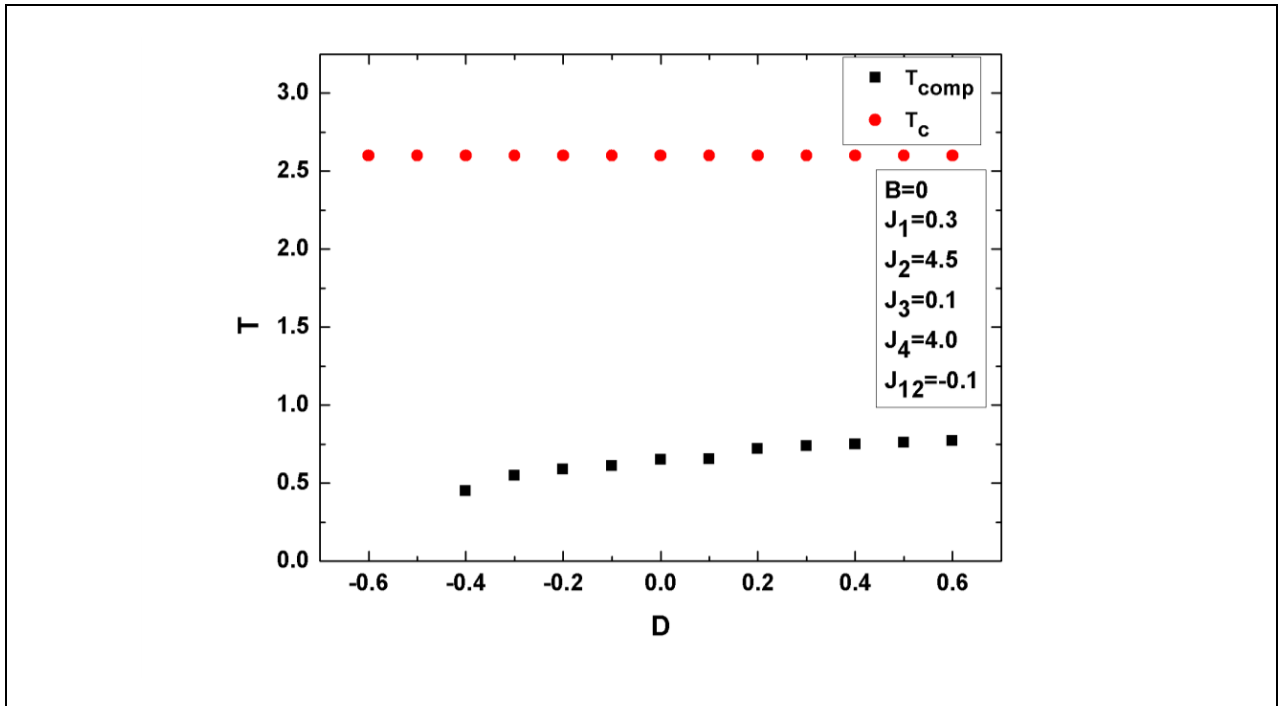

**Fig. 3** The phase diagram of the system in $(T, D)$ plane at $J_1 = 0.3$, $J_2 = 4.5$, $J_3 = 0.1$, $J_4 = 4.0$, $J_{12} = -0.1$, and $B = 0$.

*3.2 Magnetic properties*

In this subsection, we will present the general trend of the magnetization behavior, susceptibility, and the specific heat of the system as a function of temperature for selected values of the system parameters.

To illustrate the influence of the crystalline field on the specific heat, we have plotted in Fig. 4(a) the specific heat of the system as a function of temperature in the absence of the external magnetic field for different values of the crystalline field. Increasing the temperature results in increasing the internal energy of the system. At higher temperatures, the internal energy curves approach a point at which the concavity of the curves changes. This point corresponds to the critical temperature at which a second-order phase transition to a paramagnetic phase occurs. Hence, this explains the appearance of the second peak in the specific heat curves at higher temperatures.

On the other hand, decreasing the temperature results in decreasing the internal energy of the system. At lower temperatures, the internal energy curves approach a second point at which the concavity of the curves changes again because of the abrupt drop in the magnetization of sublattice-A due to weak exchange couplings $J_1$ and $J_3$. The change of the concavity in the internal energy curves at lower temperatures coincides with the location of the compensation temperature. Hence, this explains the existence of the first peak in the specific heat curves at lower temperatures. It is worth mentioning that the change of the concavity in the internal energy curves at lower temperatures is a non-critical point, i.e., no phase transition occurs at this point.

In Fig. 4(b), we have plotted the system's specific heat as a function of temperature with external applied field $B = 0.1$ and the same exchange interaction parameters as in Fig. 4(a). Again, one can remark that the location of the critical temperature shifted slightly to the right, which suggests that the critical temperature increases as the value of the external magnetic field increases. For both figures, the value of $D$ does not affect the position of the second peak. However, it affects the position of the first peak.

Fig. 5 shows the total magnetization of the system as a function of temperature for selected values of the system parameters. In Fig. 5(a), we varied the crystalline field for $J_1 = 0.3, J_2 = 4.5, J_3 = 0.1, J_4 = 4.0, J_{12} = -0.1$, and $B = 0$. The figure shows L-type magnetization. Two magnetization zero points in the magnetization curves were observed. The first one denotes the compensation temperature, whereas the second one corresponds to the critical temperature. Note that the first magnetization zero point of each magnetization curve moves left for negative values of the crystalline field, which suggests that the compensation temperature decreases as the negative value of the crystal field increases. On the other hand, increasing the crystalline field's positive value results in increasing the compensation temperature. The second magnetization zero point corresponds to the critical temperature at which a second-order phase transition to the paramagnetic phase occurs. One can note that by using large negative values of the crystal field, more magnetic domains inside sublattice-A appear even at low temperatures, decreasing the value of the total magnetization. In Fig. 5(b), we have used the same system parameters as in Fig. 5(a) but at $B = 0.1$. We have noticed that both critical and compensation temperatures have increased, attributed to the fact that the applied field tends to orient the spin along its direction.

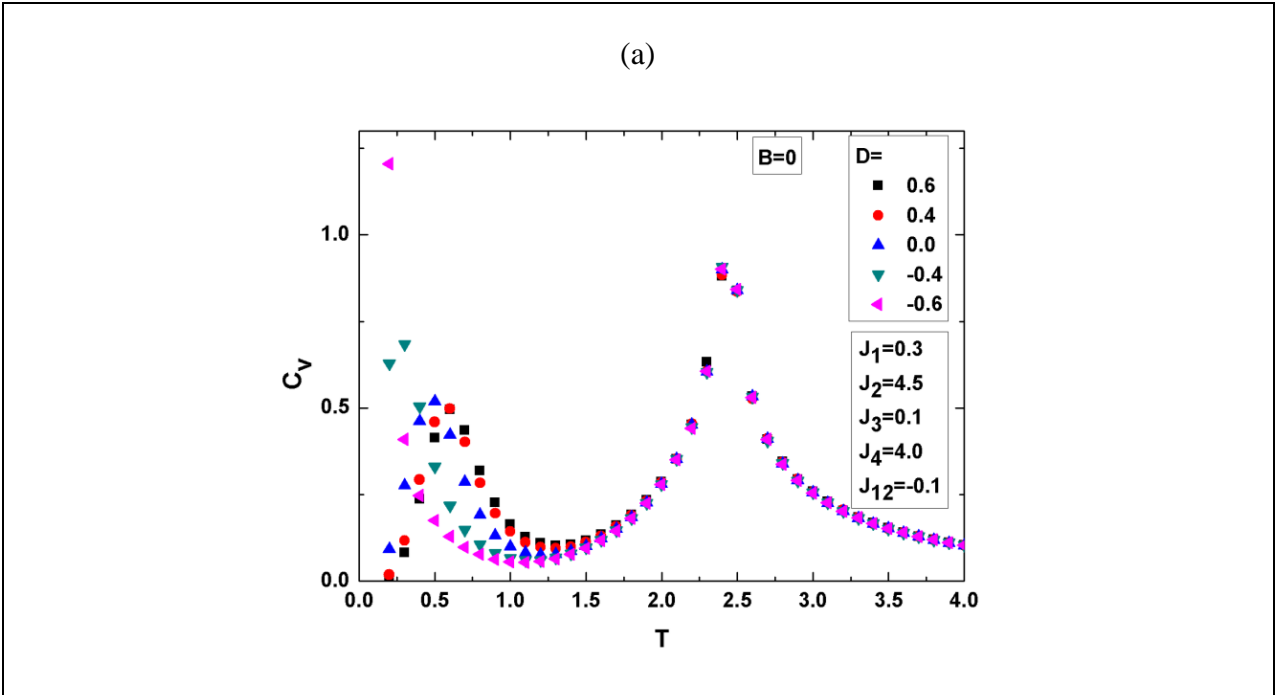

**Fig. 4** The temperature dependencies of the specific heat $C_v$ versus $T$ for different $D$ values and specific values of the exchange interaction parameters at (a) $B = 0$. (b) $B = 0.1$.

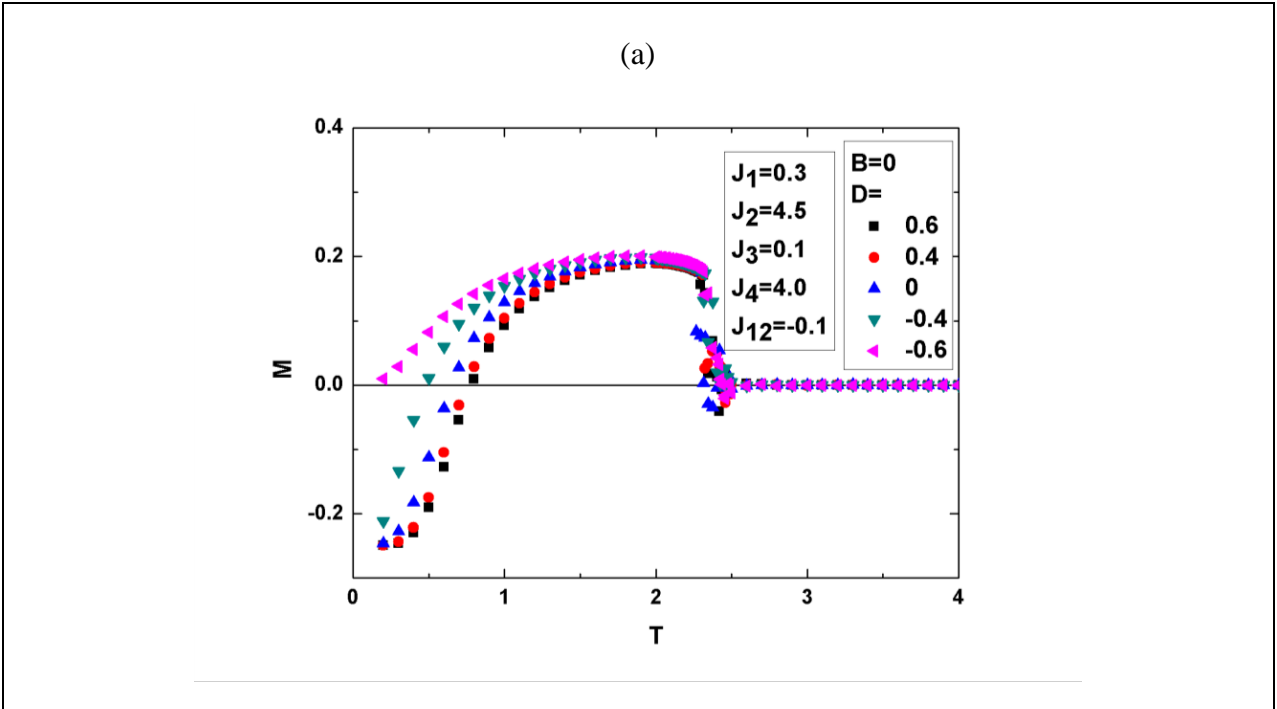

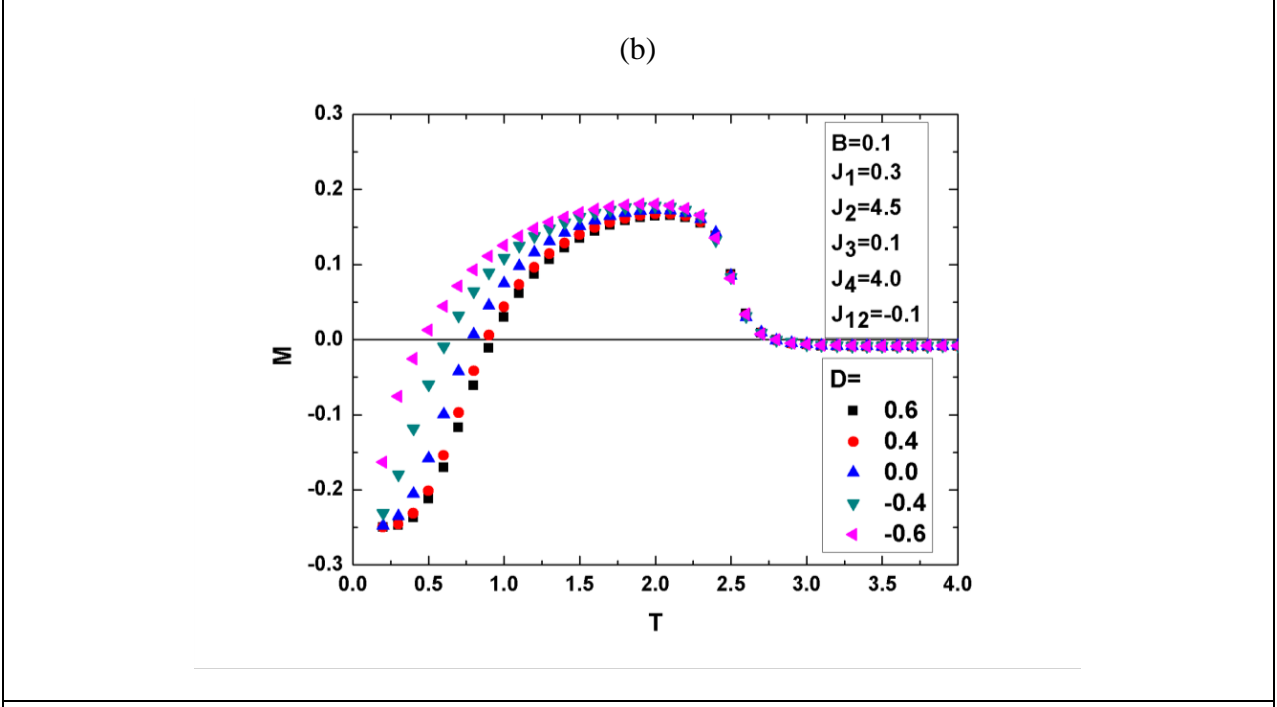

**Fig. 5 (a)** The temperature dependencies of the system's total magnetization for different $D$ values and specific values of the exchange interaction parameters at (a) $B = 0$. (b) $B = 0.1$.

To see the effect of the crystalline field $D$ on the total magnetization $M$ at a certain temperature $T$, we plot in Fig. 6 ($M$ vs. $D$) curve at $T = 1$. The total magnetization approaches zero as the crystalline field increases. This increase is attributed to the fact that using large positive values of the crystalline field makes it harder to flip the spin in sublattice-$A$, leads to decreasing the value of the total magnetization. On the other hand, the absolute value of the total magnetization increases as the hard axis crystalline field increases. The figure also shows the effect of the externally applied field. It shifts the magnetization value to be smaller. This result agreed with that found in figure (5).

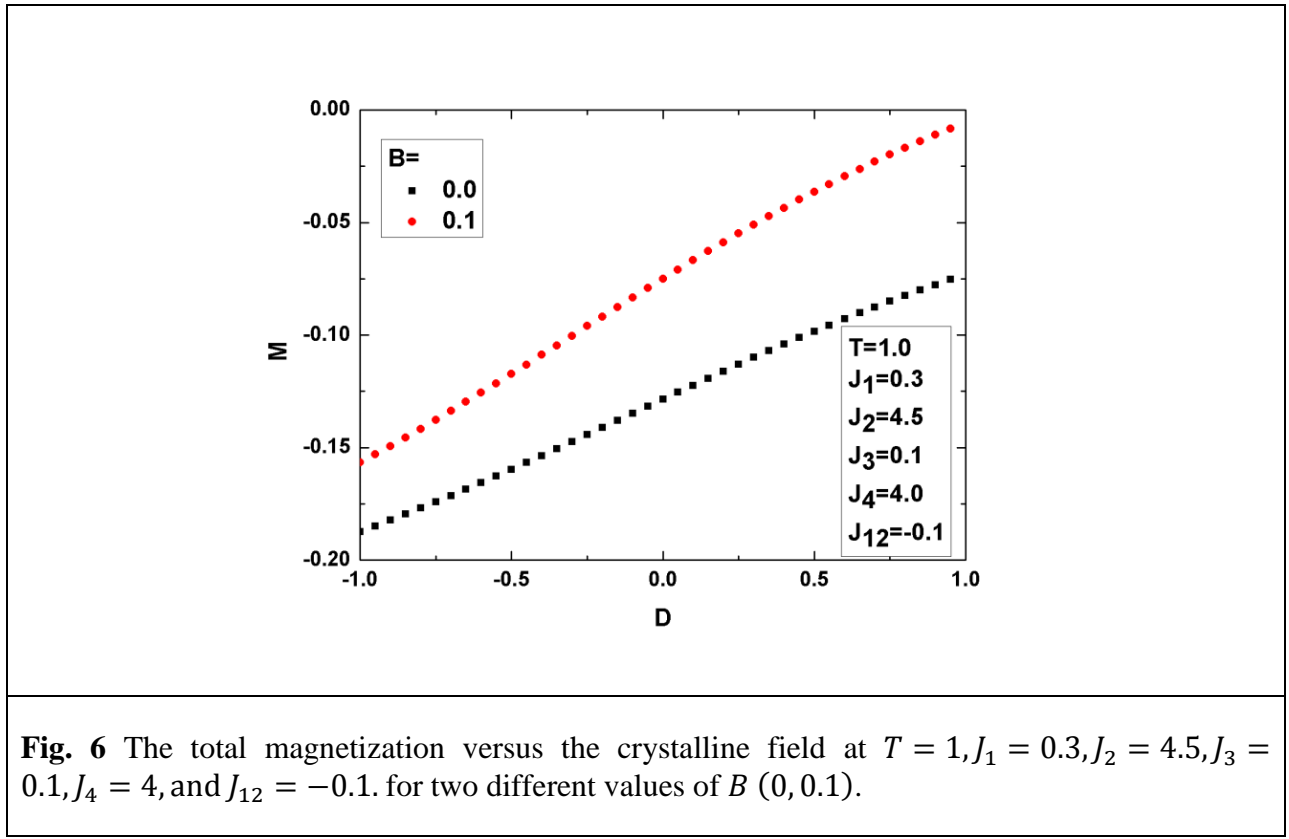

**Fig. 6** The total magnetization versus the crystalline field at $T = 1, J_1 = 0.3, J_2 = 4.5, J_3 = 0.1, J_4 = 4$, and $J_{12} = -0.1$. for two different values of $B$ $(0, 0.1)$.

The magnetic susceptibility $\chi$ versus $T$ at different values of $D$ is shown in Fig. 7(a) for $B = 0$ and Fig. 7(b) for $B = 0.1$. The figure shows double peaks where the lower peak indicates the position of $T_{comp}$ and the higher peak indicates the position of $T_C$. The crystal field effect gives higher susceptibility values at $B = 0$. The results here confirm the magnetization behavior in figure (5). Note that the location of the double peaks shifted slightly to the right in the presence of the external magnetic field. Hence, the critical and compensation temperatures increase as the strength of the magnetic field increase. The inset shows the location of the first peak of the susceptibility curves at which the compensation temperature occurs.

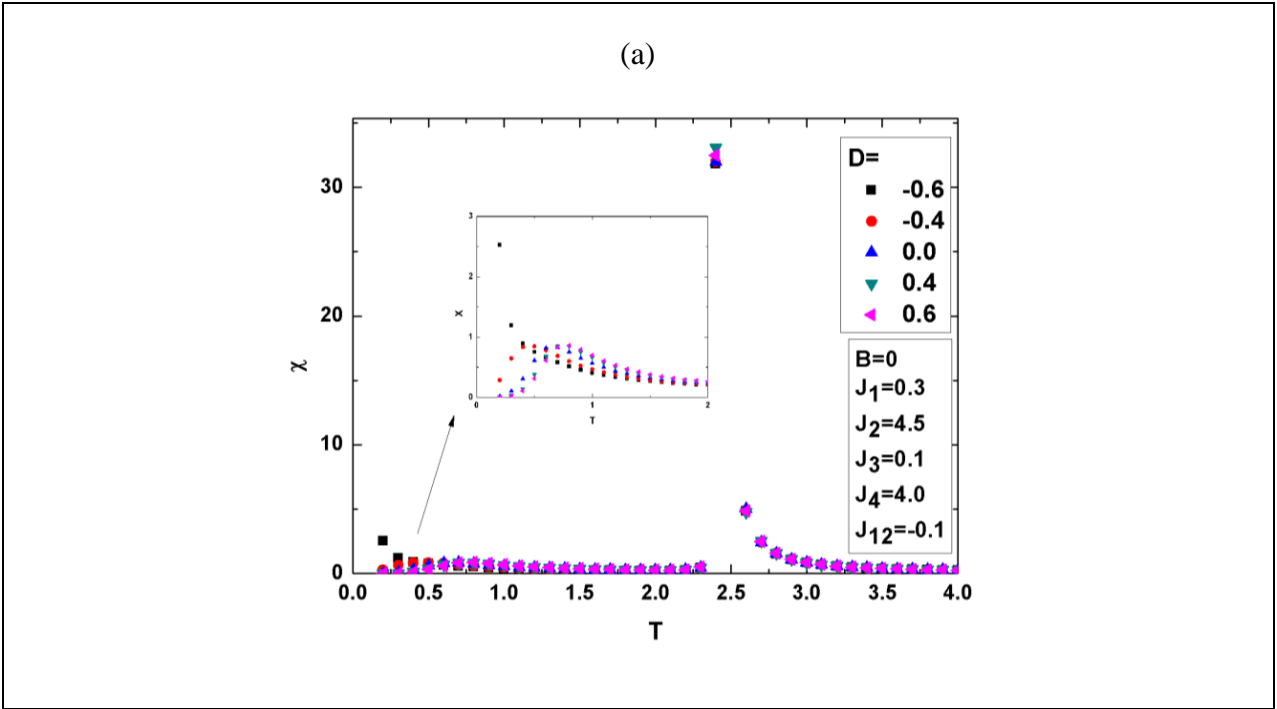

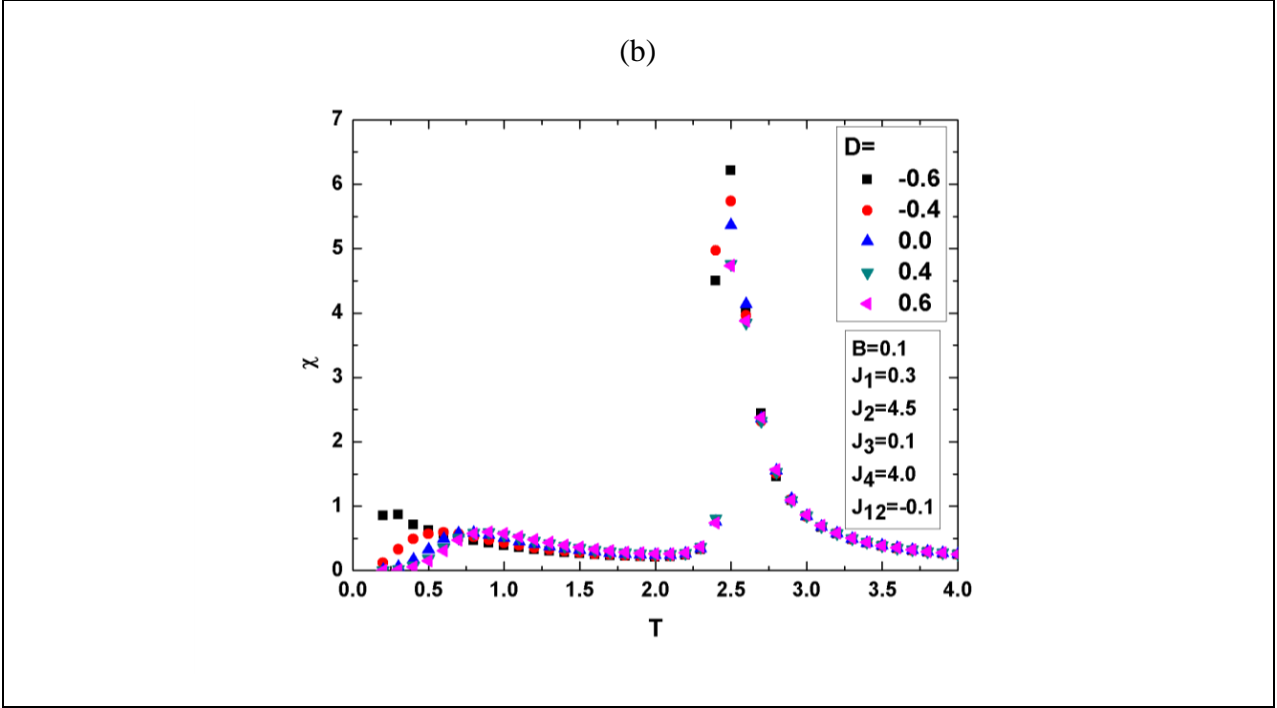

**Fig. 7** The temperature dependencies of the total susceptibilities for different values of the crystalline field at (a) $B = 0$, and (b) $B = 0.1$.

Fig. 8 shows the hysteresis loops at $T = 1$ for different values of $D$. The figure shows that the magnetization curves reach the saturation value |M|=0.75 faster for the easy axis anisotropy than hard axis anisotropy. Simultaneously, remanence magnetization and coercivity increase as $D$

decreases. Furthermore, It is worth mentioning that for the hard axis anisotropy, $D < 0$, more work is needed to reverse the direction of the magnetization, hence, the area of the loop increases. On the other hand, for the easy axis anisotropy, $D > 0$, less work is needed to reverse the magnetization direction. Hence, the area of the loop decreases.

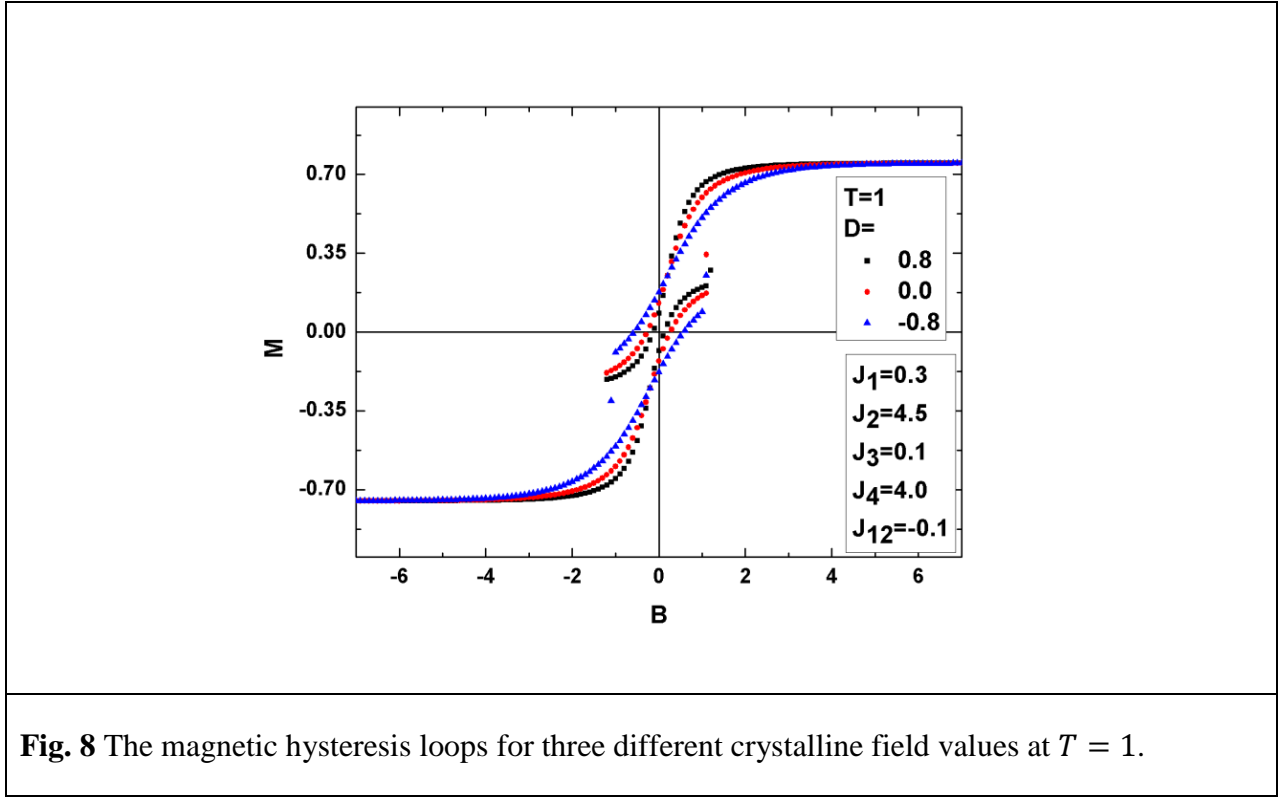

**Fig. 8** The magnetic hysteresis loops for three different crystalline field values at $T = 1$.

Fig. 9 shows the phase diagram of the antiferromagnetic coupling interaction $J_{12}$ versus the anisotropy constant $D$. The color gradient shows the value of the compensation temperature, whereas the white color in the background demonstrates the values of $J_{12}$ and $D$ at which the compensation behavior of the system disappears. For positive values of $D$, the compensation temperature increases as $|J_{12}|$ increases meaning that you need to increase the temperature to overcome the exchange interaction and flip the spin. However, for negative values of $D$, the compensation temperature could be decreasing, increasing, or increasing then decreasing as $|J_{12}|$ increase. For example, a competition between the exchange coupling parameter and the crystal anisotropy constant make the compensation temperature for $D = -1.5$ start increasing then decreasing as $|J_{12}|$ increases. The lowest compensation temperature occurs for negative values of $D$, and for these small values to occur, both values of $D$ and $J_{12}$ have to be very close to zero or at relatively high negative values of $D$ and $J_{12}$. Also, it is interesting to know that for $|J_{12}| \leq 0.6$ and $|J_{12}| \geq 0.9$ the compensation temperature increases as $D$ increases. However, for $0.6 < |J_{12}| < 0.9$ the compensation temperature decreases and then starts to increase as $D$ increases.

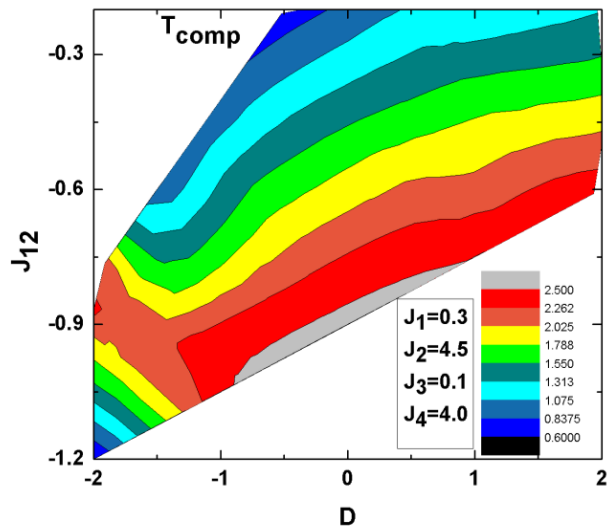

**Fig. 9** Phase diagram of the system in $(J_{12}, D)$ plane at $J_1 = 0.3, J_2 = 4.5, J_3 = 0.1, J_4 = 4.0$, and $B = 0$.

Fig. 10 shows the positive part of the hysteresis loop at $T = 1$ and different values of crystal anisotropy constant. A larger external field is needed to achieve the same degree of magnetization, $|M| = 0.75$, as the anisotropy constant decreases.

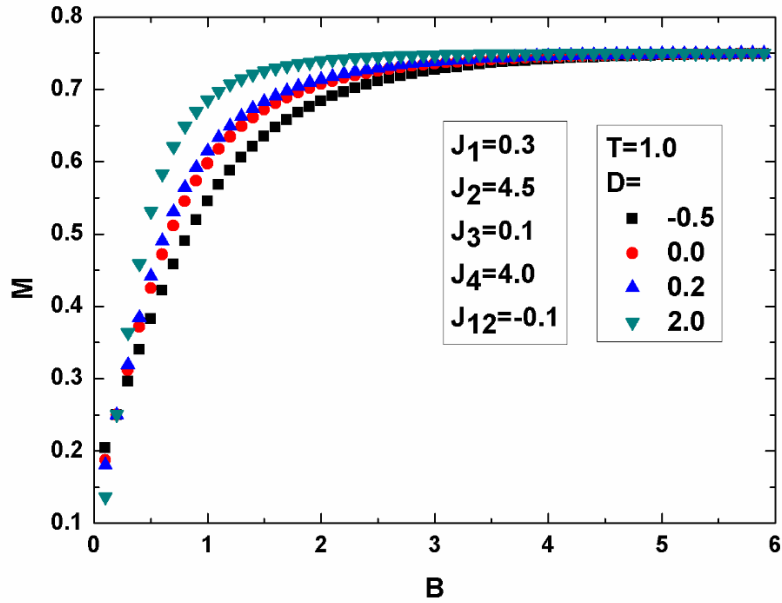

**Fig. 10** Total magnetization of the system versus the external magnetic field at $T = 1$ for $J_1 = 0.3, J_2 = 4.5, J_3 = 0.1, J_4 = 4.0, J_{12} = -0.1$ and different values of the crystal field $D$.

Fig. 11 shows the magnetization curves as a function of coupling strength $J_1$ between $S$-type atoms along [10] direction at $T = 1$ for different values of the crystalline field. The intersection points represent the value of the exchange coupling $J_1$ at which a compensation behavior in the system is achieved. For both hard and easy axis anisotropies, the value of $J_1$ should be small to obtain compensation behavior in the system. However, as $D$ increases, a smaller interaction constant $J_1$ is needed to flip the spin in sublattice $A$. For large values of $J_1$ sublattices-$A$ and $B$ are completely ordered at $T = 1$, and both hard and easy axis anisotropies have no effect on the direction of the spins in both sublattices, which explains the behavior of the total magnetization of the system $|M| = 0.25$.

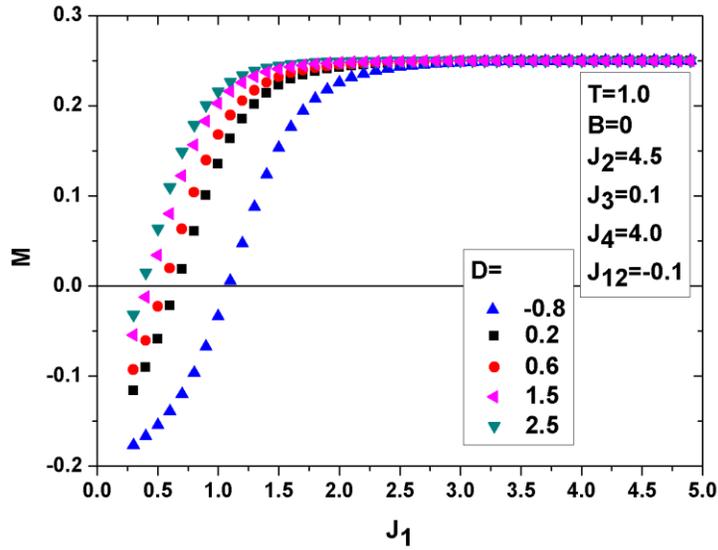

**Fig. 11** The total magnetization of the system versus the exchange coupling $J_1$ at $J_2 = 4.5$, $J_3 = 0.1$, $J_4 = 4.0$, $J_{12} = -0.1$, $B = 0$, and $T = 1.0$, for different values of the crystal field.

Fig. 12 shows the total magnetization $M$ as a function of interaction coupling strength $J_{12}$ at $T = 1$ for different values of the anisotropy constant $D$. From the figure, we noticed that for strong antiferromagnetic, $J_{12} < 0$, a magnetization value of 0.25 is achieved. This means that all spins of type $\sigma$ are $-1/2$, and type $S$ are 1, which gives a mean value of 0.25. Decreasing the value of $D$ (hard axis), one has to strengthen the coupling interaction to achieve saturation. On the other hand, if the interaction between different types of spins is ferromagnetic ($J_{12} > 1$) a constant magnetization of -0.75 is achieved. In this case, all atoms of $\sigma$ type have a value of $-1/2$, and of $S$ type have -1, which gives a mean value of -0.75. the saturation is faster along the easy axis, and you need to increase the value of $J_{12}$ to reach the saturation for negative crystalline fields. The most intriguing point is when the coupling interaction between different types of spins is zero. At this value, an absolute value of magnetization is 0.25 regardless of the anisotropy constant. The value can be achieved when the spins of different types are opposite, i.e., the system is in a ferrimagnetic phase.

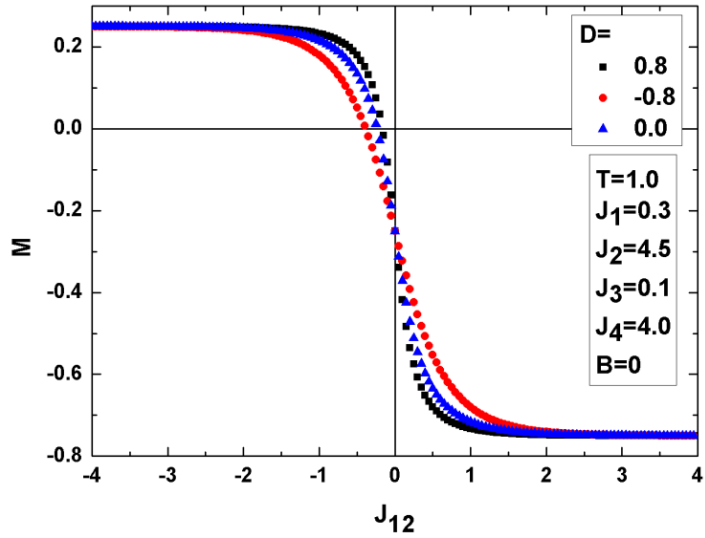

**Fig. 12** The total magnetization of the system versus the exchange coupling $J_{12}$ for $J_1 = 0.3, J_2 = 4.5, J_3 = 0.1, J_4 = 4.0, B = 0,$ and $T = 1.0$, for three different values of the crystal field.

**Conclusion**

Using Monte-Carlo simulations, we investigated the effect of the crystal anisotropy $D$ on the compensation temperature. Our results showed that an enhancement in this temperature with increasing $D$ and it appears for a wide range of $J_{12}$ values. The remanence, coercivity, and hysteresis loop area are increasing as D decreases. Large negative values of the crystal field decrease the total magnetization value as it generates more magnetic domains inside sublattice-$A$ even at low temperatures. The critical temperature and the compensation temperature increase as the applied magnetic field or $J_1$ increases. Our system has compensation temperature for small $J_1$ values and the sublattices became ordered for large $J_1$ values.

All magnetization versus temperature curves are of N-type. However, the magnetization is increased gradually with increasing $D$ (see figure 6). Moreover, the saturation of magnetization is reached faster with increasing $D$. Finally, our data reveal strong anisotropic behavior with varying $D$.